\pdfoutput=1
\documentclass[aps,pre,showpacs,twocolumn]{revtex4-1}
\usepackage{amssymb}
\usepackage{amsmath}
\usepackage{graphicx}
\usepackage{color}
\usepackage{soul}
\usepackage{hyperref}

\begin{document}

\title{From antinode clusters to node clusters: The concentration dependent transition of floaters on a standing Faraday wave}

\author{Ceyda Sanl{\i}}
\email{cedaysan@gmail.com}
\altaffiliation{Present address: CompleXity Networks, naXys, University of Namur, 5000 Namur, Belgium}

\author{Detlef Lohse}
\email{d.lohse@utwente.nl}

\author{Devaraj van der Meer}
\email{d.vandermeer@utwente.nl}
\affiliation{Physics of Fluids Group, MESA+ Institute for Nanotechnology, J. M. Burgers Centre for Fluid Dynamics, University of Twente, P.O. Box 217, 7500 AE Enschede, The Netherlands}

\date{\today}

\begin{abstract}
A hydrophilic floating sphere that is denser than water drifts to an amplitude maximum (antinode) of a surface standing wave. A few identical floaters therefore organize into antinode clusters. However, beyond a transitional value of the floater concentration $\phi$, we observe that the same spheres spontaneously accumulate at the nodal lines, completely inverting the self-organized particle pattern on the wave. From a potential energy estimate we show (i) that at low $\phi$ antinode clusters are energetically favorable over nodal ones and (ii) how this situation reverses at high $\phi$, in agreement with the experiment.
\end{abstract}



\maketitle


\section{Introduction}\label{intro} 

A small sphere floating at a water-air interface exhibits fascinating behavior when exposed to a periodic oscillation: On a standing surface wave, the floater moves either towards an amplitude maximum (antinode) or to an amplitude minimum (node). 
Whether a floater moves to the antinode or the node is determined by both the floater density relative  
to that of the carrier liquid and the floater hydrophobicity~\cite{Falkovich, FalkovichPRL, FalkovichEurPhys}: If the floater mass is larger than the displaced liquid mass the floater drifts towards the antinode, and in the reverse case it moves towards the node~\cite{surfacetension}. This drift continues throughout each wave period until the floater reaches a steady-state position, either at an antinode or at a nodal line~\cite{Falkovich, FalkovichPRL, FalkovichEurPhys}.

Thus, the dynamics of a single floater on a standing wave is quantitatively understood and node clusters of a few hydrophilic light floaters have been observed \cite{Falkovich, FalkovichPRL, FalkovichEurPhys}. On the other hand, 
the behavior of densely packed monolayers of floaters, so-called floater (or particle) rafts~\cite{DominicVella}, on a quiescent surface are shown to be dominated by 
the attractive capillary interaction among the floaters~\cite{Capillaryforce, Cheerios}. These lead to  
heterogeneity of the floater packing~\cite{MichealBerhanu}, and both granular and elastic responses of the floater raft~\cite{DominicVella}. 
In addition, the response of such a floater raft to a traveling capillary wave has been studied, in order to determine its elastic properties~\cite{PlanchetteLorenceauBiance}. 

In this 
paper, we combine the above two independent research problems into a single experiment: We study the position of hydrophilic heavy floaters on a standing Faraday wave as a function of the floater concentration $\phi$, by simply adding additional floaters to the surface. 
We experimentally show that the position of the floaters highly depends on $\phi$. For low $\phi$, our floaters accumulate at the antinodes as for the particles used in this experiment would be expected from theory~\cite{Falkovich, FalkovichEurPhys} and previous experiments~\cite{Falkovich, FalkovichPRL, FalkovichEurPhys}. Increasing $\phi$, we observe that the same hydrophilic heavy floaters cluster around the nodal lines.  
Importantly, we show that this inverted clustering is \emph{not} due to an inverted drift of a single floater, but arises as a \emph{collective effect} of many interacting floaters. 
Subsequently, we develop a potential energy estimate to explain why for high values of $\phi$ nodal clusters are energetically favored over the antinodal ones. \\
 
\section{Experiment}\label{exp} 

\begin{figure}[h!]
\begin{center}
  \includegraphics[width=8.2 cm]{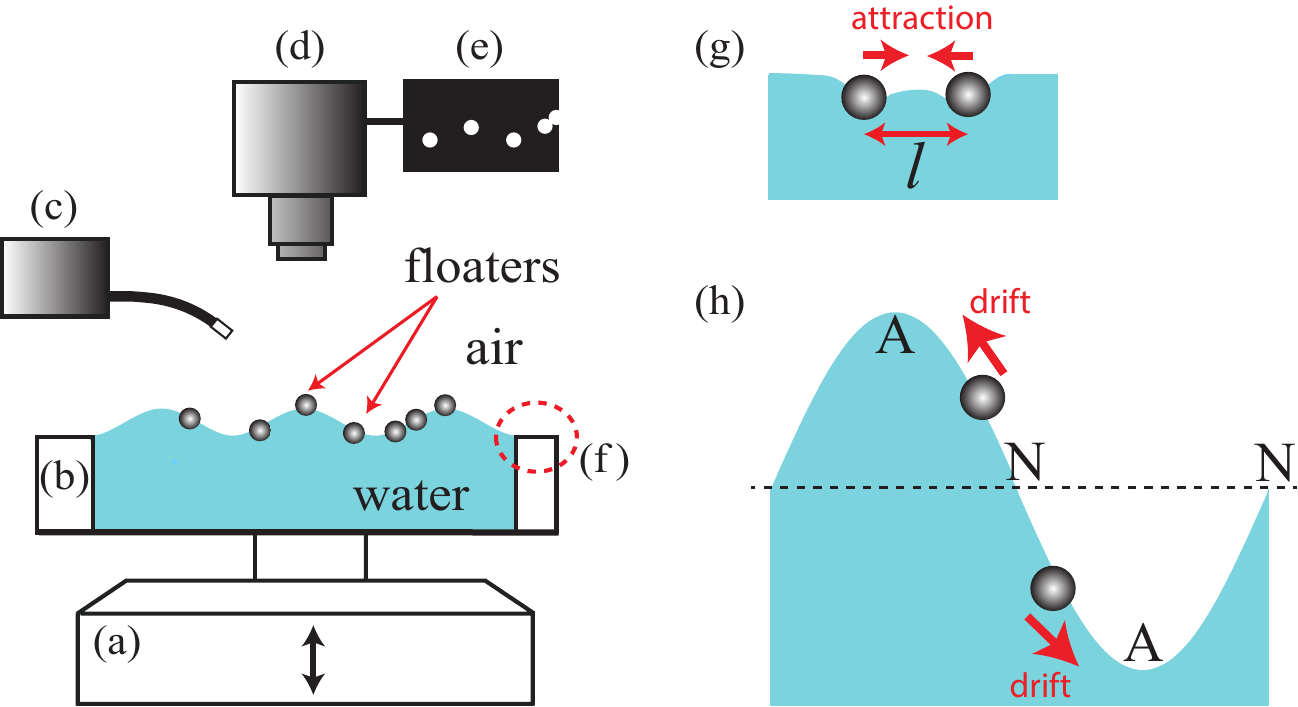}
\end{center}
\vspace{-6mm} \caption{\label{Fig1_setup} \small (Color online). Experimental setup: (a) shaker, (b) glass container, 81$\times$45$\times$10 mm$^3$, (c) Schott fiber light source, (d) Photron Fastcam SA.1, (e) an illustration of a camera image, (f) pinned brim-full boundary condition, (g) the surface deformation around our hydrophilic heavy floaters causes an attractive force, (h) the direction of the period-averaged drift of a single floater, where A and N represent the antinode and the node, respectively.}
\end{figure}

The experimental setup is illustrated in Fig. \ref{Fig1_setup}. A container, made from transparent hydrophilic glass with 10 mm height and 81$\times$45 mm$^2$ rectangular cross section is attached to a shaker. The container is completely filled with purified water (Millipore water with a resistivity $>$ 18 M$\Omega\cdot$cm) such that the water level is perfectly matched with the container edge [Fig. \ref{Fig1_setup}(f)]. Using this so-called brim-full boundary condition~\cite{Douady}, 
a static surface inclination induced by the boundary is avoided~\cite{Falkovich, FalkovichPRL, FalkovichEurPhys}. Spherical polystyrene floaters~\cite{particlesource} (contact angle $74^\circ$ and density 1050 kg/m$^3$) with average radius $R$ around 0.31 mm and a polydispersity of approximately $14\%$ are carefully distributed over the water surface to make a monolayer. To avoid surfactant effects, we clean both the container and the floaters by performing the cleaning protocol described in Ref.~\cite{cleaningprotocol}. See Appendix~\ref{sect_contact_angle} for further information on the determination of the contact angle of the floater.

A standing Faraday wave is generated using a shaker providing a vertical sinusoidal oscillation with amplitude $a_0$ and frequency $f_0$. We determine $f_0$ such that we produce a rectangular wave pattern with a wavelength in the range of $17$ to $24$ mm corresponding to frequencies ranging from $37$ to $42$ Hz (note that the standing Faraday wave frequency is equal to $f_0/2$).  
Adding floaters to the surface, we need to slightly adjust both $a_0$ and $f_0$ to obtain a well defined rectangular pattern. More details of the procedure for creating a standing Faraday wave in the presence of floaters can be found in Appendix~\ref{sectIV}.
A continuous white fiber-light source (Schott) is used to illuminate the floaters from the side as shown in Fig. \ref{Fig1_setup}(c). The two-dimensional floater positions are recorded with a high-speed camera (Photron Fastcam SA.1) at 500 frames per second. 
Each image is 546 $\times$1030 pixels (38 $\times$72 mm$^2$), which covers around $75\%$ of the total cross section area of the container. 
The vertical depth of field is taken to be large enough to capture the maximum vertical displacement ($2.5\pm0.1$ mm) of the floaters.

In the period-averaged context, there are two mechanisms that drive the floaters on the standing Faraday wave. The first one is the attractive capillary interaction~\cite{Capillaryforce, Cheerios} due to the surface deformation around the floaters [Fig. \ref{Fig1_setup}(g)], which is significant  
when the distance between the floaters $l$ is smaller than the capillary length $l_c=(\sigma/\rho g)^{1/2}$. Here, $\sigma$ is the surface tension coefficient of the interface, $\rho$ the liquid density, and $g$ the acceleration of gravity. (For an air-water interface at $20\,^\circ$C, $l_c=2.7$ mm.) The second is due to 
the standing Faraday wave, which causes  
a time-averaged drift of the floaters towards the antinodes [Fig.~\ref{Fig1_setup}(h)], which is observed and described in Refs.~\cite{Falkovich, FalkovichEurPhys}. This drift, which is discussed in greater detail in Appendix~\ref{sect_drift force}, is reminiscent of the famous Stokes drift of an object on a traveling wave.

\begin{figure}[h!]
\begin{center}
  \includegraphics[width=6.93 cm]{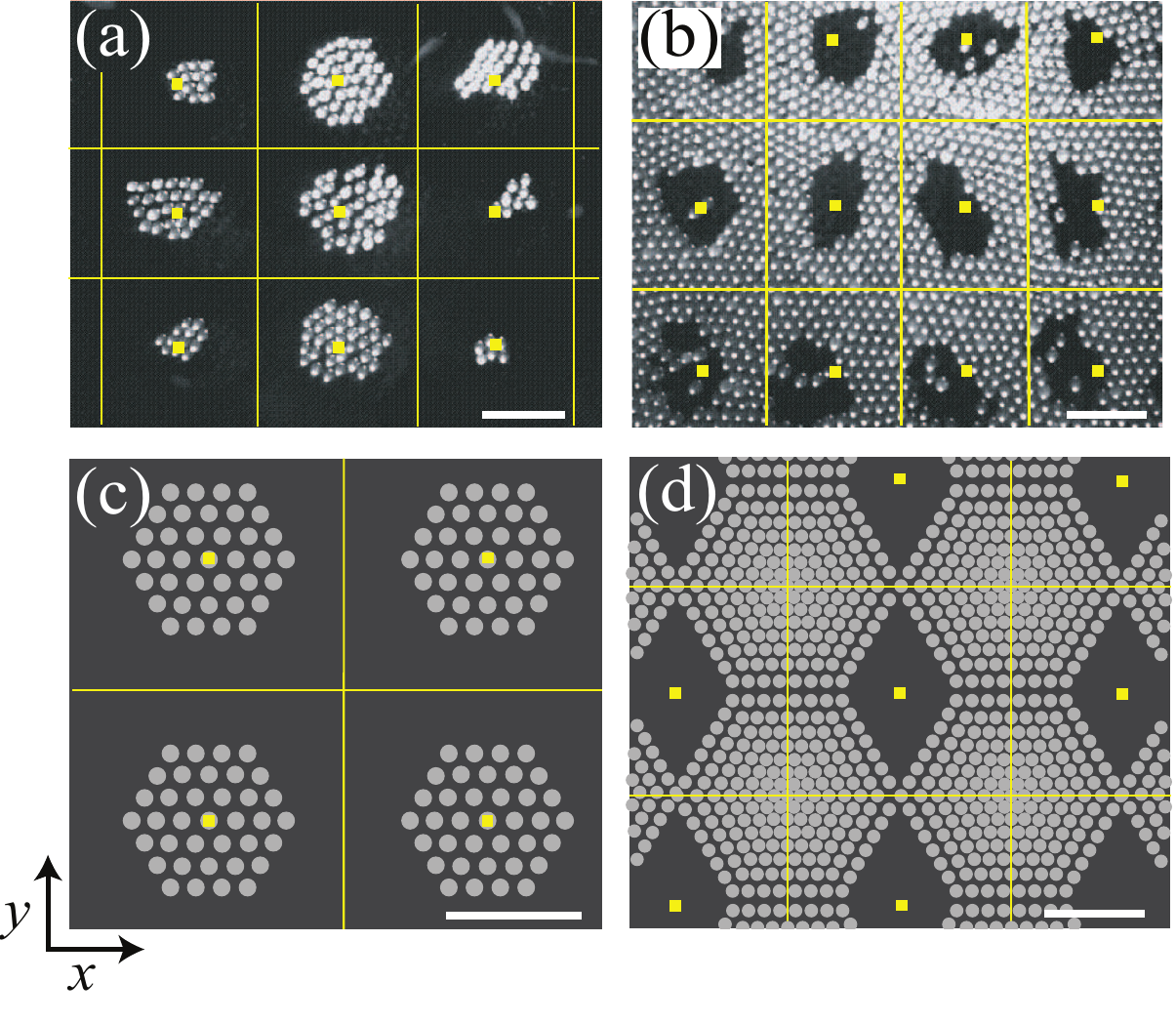}
\end{center}
\vspace{-6mm} \caption{\label{Fig2_exp_images} \small (Color online). (a), (b) Clustering of floaters on a rectangular standing wave in experiment. The snapshots show the stationary state when the surface wave elevation is nearly zero. The small yellow rectangles mark the location of the antinodes and the yellow lines that of the nodal lines. Clearly, for $\phi$ = 0.08 (a) particles cluster at the antinodes, whereas for $\phi$ = 0.61 (b) the pattern is spontaneously inverted into a large cluster around the nodal lines. Note that in (b) all particles touch whereas the average distance between particles in (a) is somewhat larger. This is due to the breathing effect explained in the text. (c,d) Artificial antinode clusters at $\phi$ = 0.10 (c) and node clusters at $\phi$ = 0.44 (d) as used in the potential energy calculation. 
The white bars indicate a length scale of 5 mm.}
\end{figure}

The control parameter of the experiment is the floater concentration $\phi$. We simply measure $\phi$ by 
dividing the area covered with floaters  
by that of the total horizontal field of view. In Fig.~\ref{Fig2_exp_images} we show a top view of the distribution of the particles in two distinct limits, namely for low $\phi$ and high $\phi$. 
The remarkable difference between the two states is clear: For low $\phi$ [Fig.~\ref{Fig2_exp_images}(a)] small clusters float around the antinodes, whereas for high $\phi$ [Fig.~\ref{Fig2_exp_images}(b)] there is one large cluster around the nodal lines. This completely inverts the pattern and the particles now seen to avoid the antinodal regions. 

To inspect this concentration-dependent clustering we introduce the correlation factor $c$, which quantifies to what extent the position of the clusters is correlated with the wave antinodes
\begin{equation}\label{c}
    c\equiv\frac{<\phi(\textbf{r},t)a(\textbf{r})>_{\textbf{r},t}}{<\phi(\textbf{r},t)>_{\textbf{r},t}}\,,
\end{equation} 
where the brackets $<>_{\textbf{r},t}$ indicate that the average is taken with respect to both space $\textbf{r}=(x,y)$ and time $t$ \cite{spatialaverage}. Here, $\phi(\textbf{r},t)$ is the floater concentration and the wave distribution $a(\textbf{r})$ is a test function that is positive at the antinodes and negative at the nodes. More specifically, 
$a(\textbf{r})$ is defined as
\begin{equation} a(\textbf{r}) = \left\{
\begin{array}{l l} 
\beta a_{cos}(\textbf{r}) &  \text{\footnotesize when $a_{cos}(\textbf{r})>0$ (antinodes),}\\
a_{cos}(\textbf{r})&  \text{\footnotesize when $a_{cos}(\textbf{r})<0$ (nodes).} \end{array} \right.
\end{equation}
Here, $a_{cos}(\textbf{r})=2\cos^2 k_xx\,\cos^2 k_yy\,-\,1$, with $k_x,k_y$ the wave numbers in the $x,y$ direction. Since with the above definition the nodal regions are three times as small as the nodal ones, a constant $\beta=3$ is introduced such that $c=0$ when the floaters are equally distributed over the two-dimensional wave surface \cite{footnote1}. 
To check the robustness of $c$ regarding the precise form of $a(\textbf{r})$, we also use a step function $a_{step}(\textbf{r})$, which 
equals $1$ at the antinodes and $-1$ at the nodes.

\begin{figure}[ht!]
\begin{center}
  \includegraphics[width=8.4 cm]{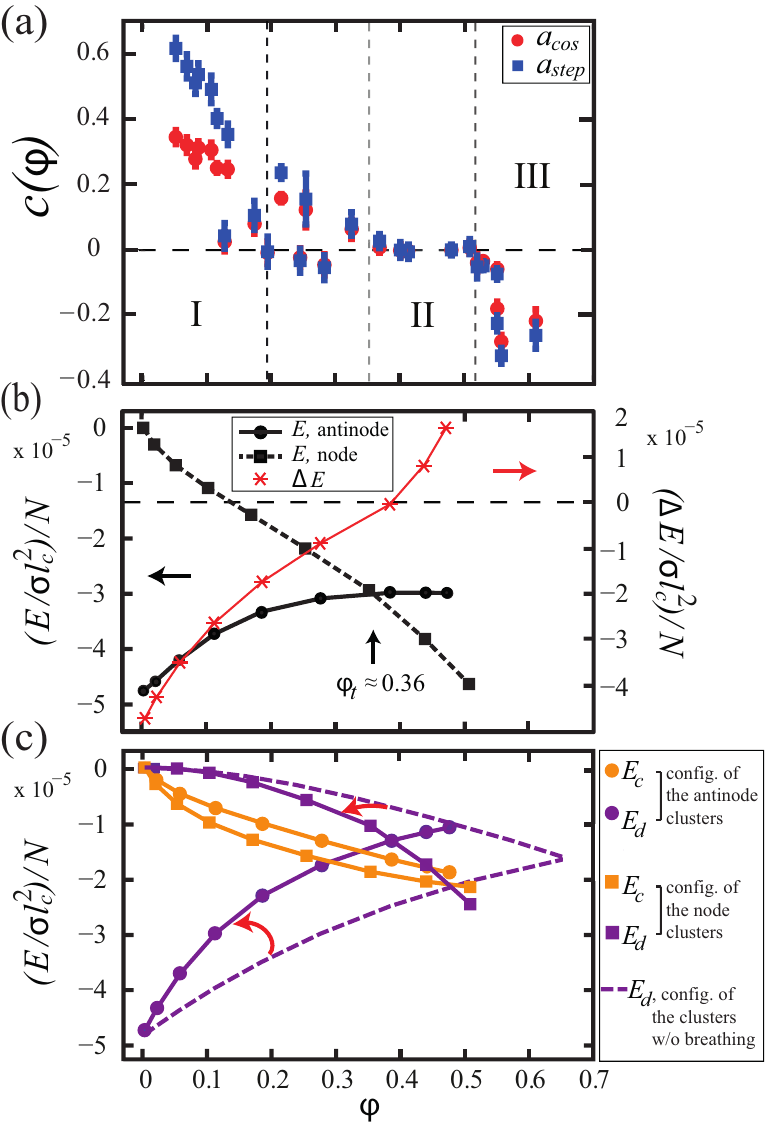}
\end{center}
\vspace{-6mm} \caption{\label{Fig3_c_Energies} \small (Color online). Experimental (a) and calculated (b), (c) transition from antinode to node clusters. 
(a) The correlation factor $c$ is plotted versus the floater concentration $\phi$ for both $a_{cos}$ (red circles) and $a_{step}$ (blue squares), where the error bars indicate the standard deviation of a single experiment. 
(b) The total potential energy $E/N$ per floater particle for the artificial patterns [see Fig.~\ref{Fig2_exp_images}(c,d)], non-dimensionalized by $\sigma l_c^2$,  
is plotted versus $\phi$ for both the antinode (black circles) and node (black squares) configurations. $\Delta E/N$ (red stars) represents the energy difference between the 
antinode and node configurations. (c) Constituents of $E/N$ versus $\phi$. Circles indicate the capillary energy $E_c/N$ (orange) and the drift energy $E_d/N$ (purple) for the antinode configurations, whereas squares indicate the same quantities for the node clusters. 
For comparison, the purple dashed lines show the drift energy $E_d$ without incorporating the breathing effect.}
\end{figure}

In Fig.~\ref{Fig3_c_Energies}(a) we present the correlation factor $c$ plotted against $\phi$ for both $a_{cos}$ and $a_{step}$. We observe three distinct regions: For low $\phi$ ($<0.2$) the clear positive value of $c$ indicates the presence of the antinode clusters (region $\textrm{I}$). Second, for very high $\phi$ ($>0.5$) we find node clusters for which $c<0$ (region $\textrm{III}$). Finally, there is a broad intermediate region $\textrm{II}$, in which we observe morphologically rich self-organized floater patterns, some of which are steadier than others. These quasi-steady patterns cause the large scatter in $c$ in the region between $\phi = 0.2$ and $0.35$. Between $\phi = 0.35$ and $0.5$, patterns are quite dynamic leading to an even spreading of particles over the waves ($c \approx 0$).  

In addition to the position, another remarkable difference between the antinode and the node clusters is hidden in their dynamics during a single wave period: Experimentally we observe that in the antinode clusters the floaters periodically move away from and towards the antinode [Fig. \ref{Fig4_designedclusters}(a)]. This happens because when the wave reaches its maximum the (downward moving) floaters move away from the antinode, whereas in the minimum they move towards  it. We call this periodic motion at the antinode clusters \emph{breathing}. In contrast, nodal clusters do not breathe; instead the clusters as a whole oscillate back and forth around the nodal lines [Fig. \ref{Fig4_designedclusters}(b)]. As a result, the floaters in the node clusters stay closely together without changing their relative distance (which is approximately equal to the particle diameter $2R$), whereas the period-averaged distance between the particles in the antinode cluster is significantly larger than $2R$~\cite{SupplementaryMovie}.  The breathing phenomenon is discussed in more detail in Appendix~\ref{sect_drift force}.


\section{Potential energy estimate}\label{est} 

\begin{figure}[ht!]
\begin{center}
  \includegraphics[width=8.2 cm]{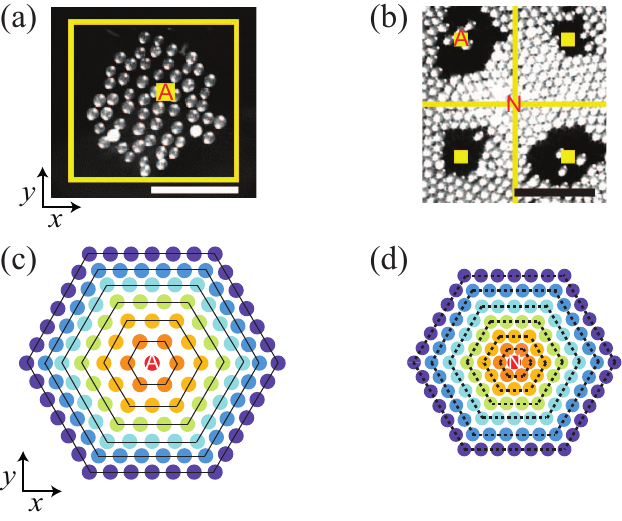}
\end{center} \vspace{-6mm} \caption{\label{Fig4_designedclusters} \small (Color online). The breathing effect: When we compare an experimental antinode (a) with a node cluster (b), we clearly see that particles in the first are much farther apart due to breathing (see text). Again, the antinodes (A) are marked by small yellow rectangles and the nodes (N) with yellow lines. The bars indicate a length scale of 5 mm. We artificially design hexagonal clusters  to incorporate this breathing effect: An antinode cluster (c) is grown by adding hexagonal rings at decreasing increments $r_{nn}$ starting from a large initial value, whereas a node cluster (d) is grown from a close-packed hexagonal structure with increasing increments $r_{nn}$. The color coding identifies consecutive rings.}
\end{figure}

Now, what is the reason for the observed pattern inversion? To answer this question we estimate the energy in artificially created node and antinode clusters, which are inspired by our experimental observations [Figs. \ref{Fig4_designedclusters}(a), \ref{Fig4_designedclusters}(b)]: 
The antinode cluster is modeled as a two-dimensional static hexagonally packed cluster where the distance between the neighboring floaters increases towards the antinode point (A) [Fig. \ref{Fig4_designedclusters}(c)] to implement the observed breathing effect. The distance here can be considered as the period-averaged experimental distance between the floaters. The node cluster, in contrast, is designed as a two-dimensional hexagonal cluster where the distance between the neighboring floaters sitting exactly at the crossing of two nodal lines (N) is equal to an average floater diameter $2R$ [Fig. \ref{Fig4_designedclusters}(d)]. Furthermore, the distance slightly increases away from N. Further details on the artificial antinode and node cluster configurations can be found in Appendix~\ref{sectII}.

During the motion of the floaters on the wave there is an intricate exchange of wave energy (input), potential energy, kinetic energy, and dissipation (output). However, in a steady state the input and output must balance and since the particles return to (approximately) the same positions after each period of the wave it is sufficient to compare the potential energy $E$ of the floaters for the two situations. This potential energy has two contributions, due to the drift and due to the capillary attraction.  

The first contribution to $E$ is the capillary energy $E_c$, which we estimate as the sum of the capillary energies of each floater pair $E_c(l_{i,j})$, where $l_{i,j}$ is the 
distance between floaters $i$ and $j$. Here, we use the approximation $E_c(l_{i,j}) = A_c K_0(l_{i,j}/l_c)$, where $K_0$ is the zeroth order modified Bessel function of the second kind. This approximation is valid for small surface deformations, i.e., for small spheres, loosely packed structures or relatively distant spheres \cite{Capillaryforce, Cheerios}. [Both the size and the density of a sphere are important in judging whether the linear  
approximation is applicable. To this end, we check the Bond number $B$ for our spheres and find that $B\ll1$, i.e., the approximation is valid (see Appendix~\ref{sect_contact_angle}).] Studies that comparatively discuss the exact solution of the capillary force of floaters of similar size suggest that the difference with the approximation  
is less than $2\%$~\cite{Vassileva}.
The second contribution to $E$ is the drift energy $E_d$. It is the sum over the single-floater drift energy $E_d(x_i,y_i) = A_d (1-\cos 2k_x x_i) (1-\cos 2k_y y_i)$, where $(x_i,y_i)$ is the position of floater $i$. Note that the prefactors $A_c$ and $A_d$ are known functions of particle, liquid, and wave properties. The full expressions for $E_c(l_{i,j})$ and $E_d(x_i,y_i)$, including prefactors, are provided in Appendix~\ref{sectIII}.

Subsequently, we use the above expressions to estimate the potential energy $E$ in our antinode and node cluster configurations [Figs. \ref{Fig2_exp_images}(c), (d)] as a function of the floater concentration $\phi$ (i.e., the number of particles $N$) and compare them in Fig.~\ref{Fig3_c_Energies}(b).  
For increasing $\phi$, the energy per floater $E/N$ increases for the antinode clusters, whereas it decreases for the node clusters. As a result, there is a crossover $\phi_t\approx0.36$ separating a low $\phi$ region, where the antinode clusters are energetically favorable, from a high $\phi$ one, where the node clusters have lower potential energy. In addition, $\phi_t$ lies in the transition region of Fig.~\ref{Fig3_c_Energies}(a) and is therefore in agreement with the experiment.

To examine the physical reason for this crossover,  in Fig.~\ref{Fig3_c_Energies}(c) we  
turn to the constituents of $E$, namely $E_c$ and $E_d$. 
For the capillary energy $E_c$ there is hardly any difference between the node and antinode clusters, except for a slightly milder decrease for the latter, caused by the larger average distance between the floaters due to the breathing. 

Things are very different for the drift term: For small $\phi$ the node clusters initially have a high drift energy $E_d/N$ per floater and the antinode clusters are favorable. When we increase $\phi$ without including the breathing effect, i.e., both clusters are just hexagonally packed with nearest-neighbor distance $2R$, the energy per floater in the node clusters decreases and that of the antinode clusters increases until they meet for a very high value of $\phi$, corresponding to an almost completely floater-covered surface [dashed lines in Fig.~\ref{Fig3_c_Energies}(c)]. However, when we do include the breathing effect in our calculation, $E_d/N$ increases much faster for the antinode cluster due to the large average distance of the particles near the antinodes. Similarly, $E_d/N$ increases somewhat more rapidly for the nodal clusters. The result is that the crossover shifts to a moderate value of $\phi$, namely $\phi_t \approx 0.36$. This implies that nodal clusters now already become energetically favorable when the surface is not yet covered with particles, which causes the inverted patterns to exist.\\


\section{Conclusion}\label{conc} 

In summary, in this paper 
we study the role of the floater concentration $\phi$ on the spatial distribution of macroscopic spheres floating on a standing Faraday wave. For low $\phi$, we experimentally observe that 
hydrophilic heavy floaters form clusters at the antinodes, suggested by the theory \cite{Falkovich, FalkovichEurPhys}. For high $\phi$, the same floaters unexpectedly self-organize into the \emph{inverse} pattern, namely a large cluster around the nodal lines of the wave. To understand such a collective behavior, we calculate the potential energy of the floater system and are able to  
explain our observations in both limits. More specifically, the transition point $\phi_t$ obtained from our energy calculation lies within the experimental transition region.

We find that the observed breathing effect is essential for the existence of the crossover. The breathing creates a significant difference in the drift energy such that the node clusters are energetically favorable already when only drift energy is taken into account. The role of the capillary interaction is just to keep the floater particles self-organized in rafts; without this attractive interaction the floaters would be freely drifting around instead of forming clusters. 

Whereas our potential energy argument nicely accounts for the existence of the stable antinode and node patterns, it is not able to capture the large transitional region that was observed between $\phi=0.2$ and $0.5$. Presumably, what happens in this region is that the antinode clusters become too large to stay pinned at the antinode regions and start to wander into the nodal regions under the influence of the wave motion. Characterizing these patterns will be the subject of future research.

The work is part of the research program of FOM, which is financially supported by NWO. 

\appendix

\section{Floater details, drift force, and breathing}\label{sectI}

In this series of Appendices  
some technical details about both the experiments and the calculations presented in the main text of the article are provided. 
In Appendix \ref{sectI}, we discuss the contact angle calculation of the floaters, the direction of the corresponding wave drift, and further details of the breathing motion. In Appendix \ref{sectII}, the procedure to construct two-dimensional clusters on a standing wave is presented. Subsequently, in Appendix \ref{sectIII}, we provide both the capillary and the drift energies used in the main text of the article 
but now together with the prefactors. Finally, the experimental details of creating a standing Faraday wave with the floaters are described in Appendix \ref{sectIV}.

\subsection{Calculation of the floater contact angle}\label{sect_contact_angle}

\begin{figure}[h]
\begin{center}
  \includegraphics[width=0.7\linewidth]{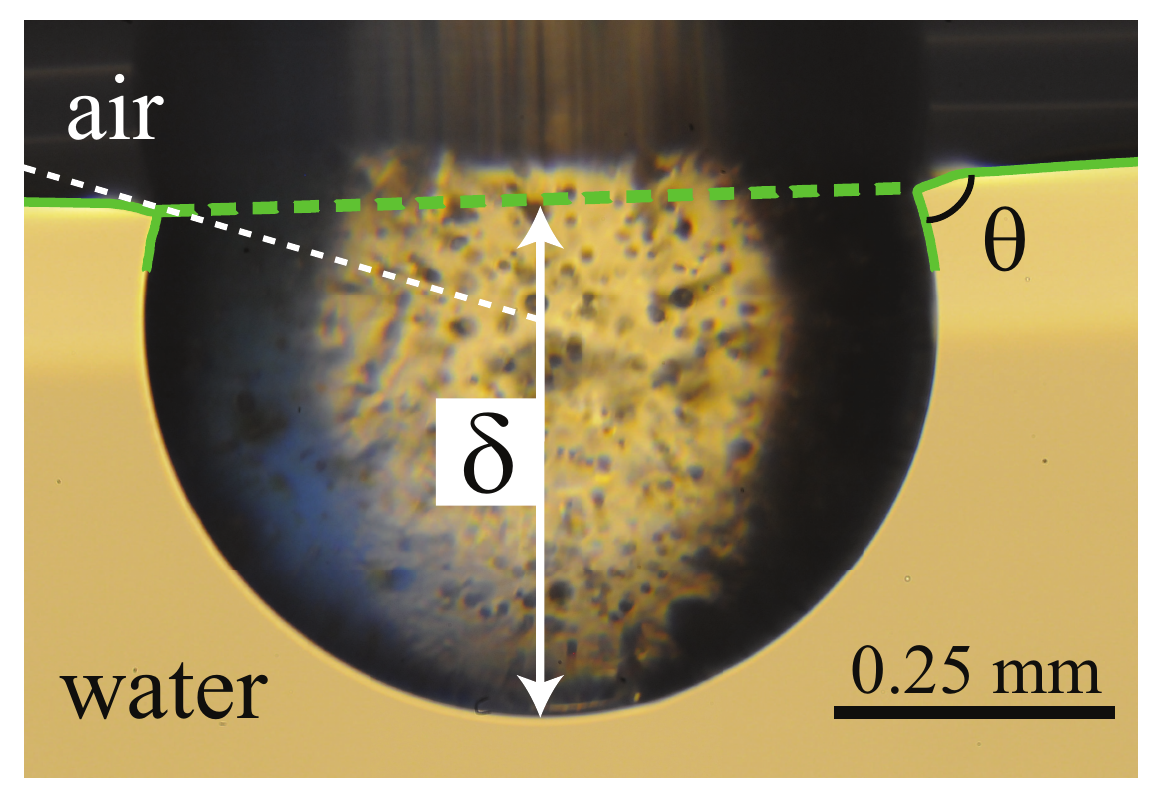}
\end{center}
\vspace{-6mm} \caption{\label{suppl_Fig1_floaterinterface}\small (Color online) A polystyrene sphere floating in a static equilibrium at an air-water interface is imaged from the side. The solid green line represents the interface. The green dashed line displays the circular contact line around the floater. The white dashed line indicates the surface normal of the contact line. The contact angle $\theta$ and the depth of the submerged part $\delta$ are shown when the floater is at the vertical force balance: The surface tension acts upwards to satisfy the balance since $M>m_d$.}
\end{figure}

Here, we calculate the contact angle of the floaters that are used in the experiment based on the static force balance from Ref.~\cite{Cheerios}. When a cleaned polystyrene spherical floater [see Ref.~\cite{cleaningprotocol} for the cleaning protocol] with density $\rho_s$ is put on an air-water interface, the water surface is deformed to satisfy the vertical force balance: The sum of the weight of the sphere, the buoyancy force and the surface tension force should be zero.
The surface deformation due to a single floater in a static equilibrium is shown in Fig.~\ref{suppl_Fig1_floaterinterface}, where $\theta$ is the contact angle and $\delta$ is the depth of the submerged part.

In this situation, the weight of the sphere, $Mg$, where $M$ is the mass of the floater and $g$ is the acceleration of the gravity, is larger than the buoyancy force, $m_d\,g$, where $m_d$ is the displaced mass, so that the surface tension force acts upwards. The fact that the contact angle $\theta$ is  
smaller than $90^\circ$ indicates that the floater is hydrophilic. The interface cannot be photographed well enough to accurately determine $\theta$ by  image analysis. To determine $\theta$ nevertheless, the expression for $\delta$ derived in Ref.~\cite{Cheerios} using the vertical force balance is employed
\begin{equation}\label{delta}
    \delta\approx R [1+\cos\theta+B\,\Sigma(\theta,D)],
\end{equation} where $R$ is the floater radius, and $B$ is the Bond number, $B=(\rho_w-\rho_a)gR^2/\sigma$. Here, $\sigma$ is the surface tension coefficient of the water-air interface, and $\rho_w$ and $\rho_a$ are the densities of water and air, respectively. $\Sigma(\theta,D)$, which is a function of $\theta$ and $D$, $D=(\rho_s-\rho_{a})/(\rho_w-\rho_a)$, is given by~\cite{Cheerios}
\begin{equation}\label{Sigma}
   \Sigma(\theta,D)=\frac{2D-1}{3}-\frac{\cos\theta}{2}+\frac{\cos^3\theta}{6}.
\end{equation} The expression given by Eq. (\ref{delta}) is a leading order approximation in 
$B$ and is valid for small surface deformations, when $B \ll 1$. In our case, $B=0.0091$ and the calculation gives $\theta \approx 74.3^\circ$, where 
the physical properties  
of air and water are taken to be the standard values at  
a temperature of $20\,^\circ$C. This is consistent with the directly observed contact angle.

\subsection{Drift force and breathing}\label{sect_drift force}

We now provide the theory of the (time-averaged) motion of the floater on the standing wave~\cite{Falkovich, FalkovichPRL, FalkovichEurPhys} and discuss the agreement with our experimental result when the floater concentration $\phi$ is low. We further connect the time-resolved floater motion to the breathing motion introduced in the main text, Section~\ref{exp}.

\emph{Drift} $-$ Firstly, in Refs.~\cite{Falkovich, FalkovichEurPhys}, the drift force is derived for a spherical particle with a given  
contact angle floating on a one-dimensional wave, which --averaged over a single wave period-- is equal to 
\begin{equation}\label{drift_ave_force}
    f(x)=\int_0^{2\pi/\omega_k} f(x,t)\,dt=\frac{1}{4} k\,a^2\omega^2_k\left(M-m_d\right)\sin2kx,
\end{equation} where $\omega_k$ is the angular frequency of the wave, $k$ is the wave number, $a$ is the wave amplitude, $M$ is the mass of the floater, and $m_d$ is the mass of the displaced liquid \cite{Afootnote}. From Eq. (\ref{drift_ave_force}) it can be understood that the direction of the drift force for a single floater depends on $M-m_d$: If $M-m_d>0$, the drift is towards the antinodes, otherwise the drift is towards the nodes. For the particles used in our experiment, $M-m_d>0$ as discussed in Appendix \ref{sect_contact_angle}. Therefore, in our case, the drift force is directed towards the antinodes, which is consistent with our experimental observations at low $\phi$ (see Fig. \ref{Fig3_c_Energies}(a) 
and Ref. \cite{SupplementaryMovie}).

Furthermore, the magnitude of the time averaged drift force depends on 
$k$ and $\omega$, both of which are varied only very slightly in our experiments, and on the squared amplitude $a^2$. Although we increase the amplitude $a_0$ of the shaker significantly this is only done to keep the amplitude $a$ of the standing wave as constant as possible (see Appendix~\ref{Frequency and amplitude of a standing Faraday wave}). Note that, considering the dependence on the position $x$, no drift is experienced if the floater 
sits either at the antinodes or at the nodes and that the drift is maximum when the floater is positioned between the antinode and the node.\\

The calculation of the drift force is a period-averaged calculation. 
There is, however, interesting dynamics hidden in a single wave period.  
Now, we will try to elucidate the time-resolved motion of the sphere on a standing wave using qualitative arguments, without turning to the full equations as was done in Refs.~\cite{Falkovich, FalkovichEurPhys}.

\begin{figure}[h]
\begin{center}
  \includegraphics[width=1.0\linewidth]{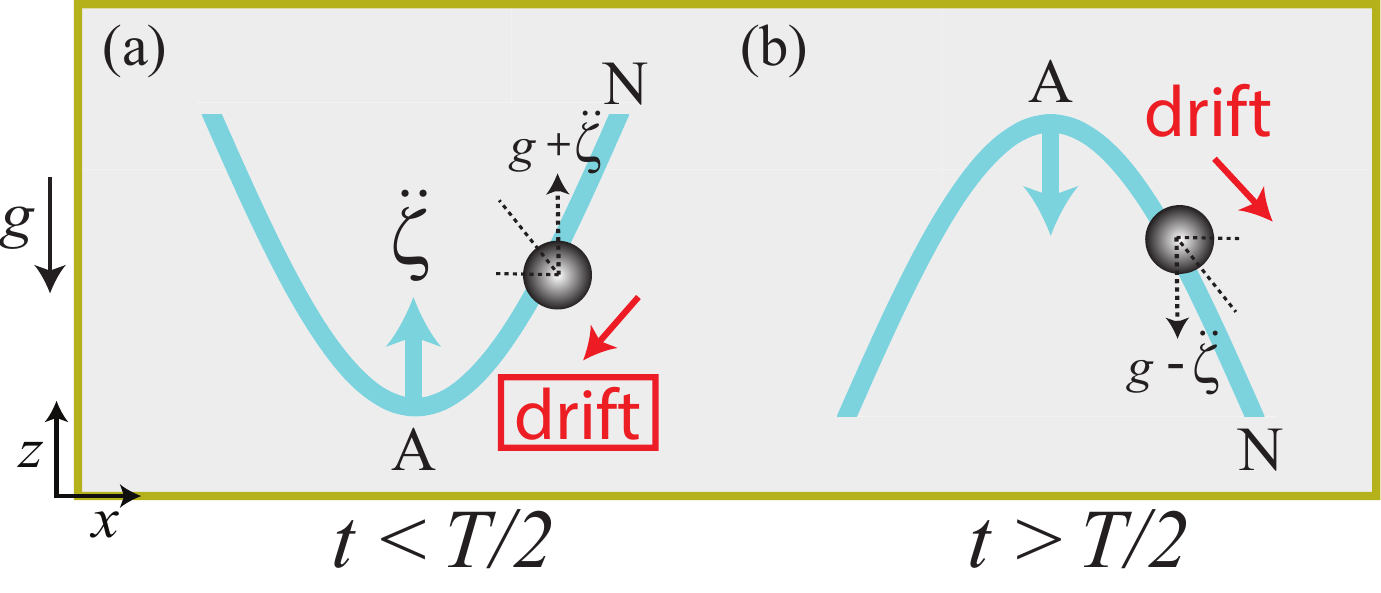}
\end{center}
\vspace{-6mm} \caption{\label{suppl_Fig2_floaterdynamics}\small (Color online) The wave elevator: The asymmetry in the vertical floater acceleration and the corresponding drift are illustrated for a sphere with $M>m_d$. $\ddot{\zeta} \approx \partial^2\zeta/\partial t^2$ is the vertical surface wave acceleration and $T$ is the wave period. 
Since the contribution of (a) [$t<T/2$] is larger than that of (b) [$t>T/2$], on average, the floater drifts towards the antinode (A).}
\end{figure}

Let us consider a small sphere with mass $M$ floating at a curved interface, which we imagine to be static. If $M$ is larger than the displaced mass $m_d$ --as is the case for the floaters discussed here-- there is an unbalanced excess vertical force driving the floater towards a local minimum. For an oscillating curved interface, such as the surface of our standing wave, the location of minima and maxima vary within a wave period $T$. When $t<T/2$ [Fig. \ref{suppl_Fig2_floaterdynamics}(a)], there is a local minimum at the wave antinode (A), whereas in the second half of the period ($t>T/2$) it represents a local maximum [Fig. \ref{suppl_Fig2_floaterdynamics}(b)]. Consequently, our single floater moves towards A for $t<T/2$ and moves towards N for $t>T/2$. Now, what does this imply for the drift the floater experiences?

The vertical wave acceleration $\ddot{\zeta}\,\widehat{\textbf{z}}$ oscillates with respect to $t$. (Here, we use $\ddot{\zeta}=\partial^2\zeta/\partial t^2$ neglecting the convective terms for simplicity.) The vertical acceleration which the floater experiences is $g+|\ddot{\zeta}|$ when $t<T/2$ and $g-|\ddot{\zeta}|$ when $t>T/2$ [Fig. \ref{suppl_Fig2_floaterdynamics}]. Since the floater acceleration is larger for $t<T/2$ the contribution of this part of the wave cycle to the drift is larger. Therefore, in the time-averaged situation, the sphere drifts towards A, consistent with 
Eq. (\ref{drift_ave_force}) \cite{Afootnote2}. The mechanism discussed here resembles an accelerating elevator, and can therefore be called a wave elevator.

\emph{Breathing} $-$ In addition to predicting the direction of the drift, the argument from Fig.~\ref{suppl_Fig2_floaterdynamics} also provides us with a qualitative picture of how the floaters move on top of the Faraday wave: In the first half of the period of the wave particles move towards one of the antinodes that attain their minimum in this half and in the second half the floaters move away from them, i.e., towards the antinodes that have their minimum in the second half of the wave period. Whereas single particles just wiggle back and forth in this manner, this has large implications for the motion of a cluster of particles.

\begin{figure}[h]
\begin{center}
  \includegraphics[width=1.0\linewidth]{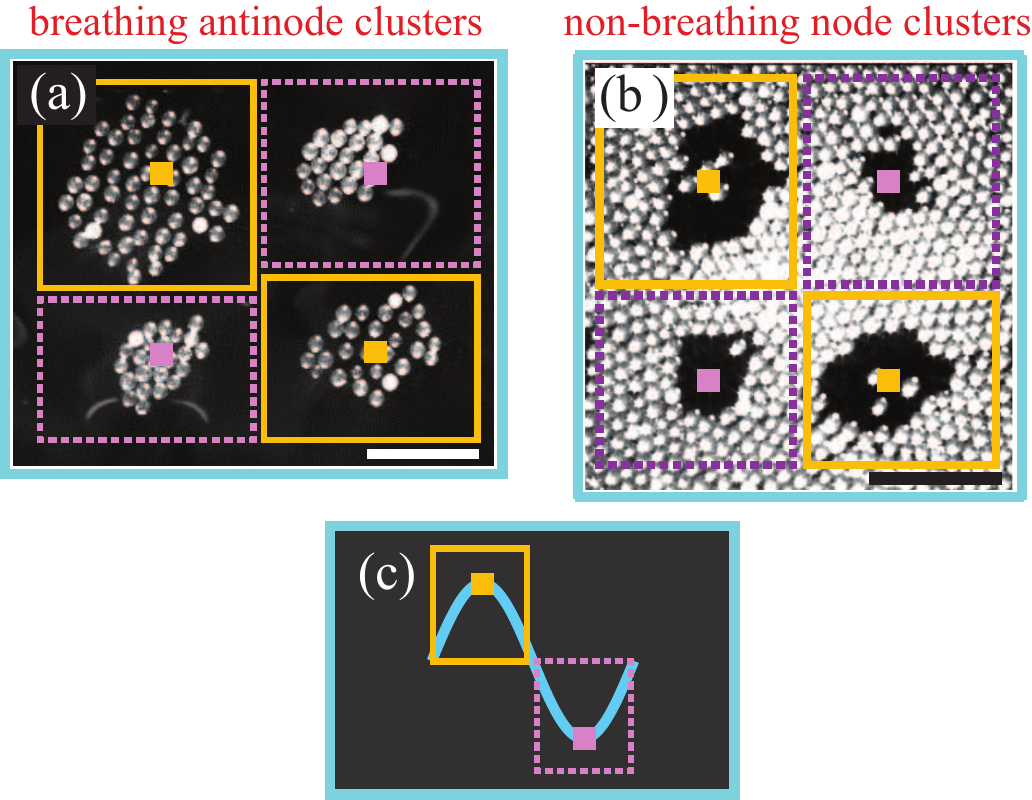}
\end{center}
\vspace{-6mm} \caption{\label{suppl_Fig3_breathing}\small (Color online) 
(a) Breathing antinode clusters are observed for low $\phi$ in the experiment. Note that in this snapshot the upper-left and lower-right antinodes are in their wave maximum (orange solid rectangles) whereas the other two are in their minimum (purple dashed rectangles). (b) Non-breathing node clusters are found for high $\phi$. Also here the upper-left and lower-right antinodes are in their wave maximum. Clearly all neighboring floaters are at the same relative distance, namely the particle diameter. (c) Side view of a one-dimensional standing wave, with the maximum (downward acceleration) indicated by the orange solid rectangle and the minimum (upward acceleration ) by the purple dashed rectangle. 
The bars indicate a length scale of 5 mm.
}
\end{figure}

For a cluster of particles that is located around a nodal line, neighboring particles move in the \emph{same} direction and therefore the cluster just oscillates back and forth as a whole. The capillary attraction between the floaters will keep the cluster together, and particles will typically touch [Fig. \ref{suppl_Fig3_breathing}(b)]. Therefore, the distance between the floaters does not vary within a wave period and is around the average floater diameter. Things are different for a cluster of particles around an antinode. Here particles are pushed towards the antinode (which is now a minimum) during the first half period and driven away from that point (now a maximum) during the second half: Antinode clusters breathe. When the floaters move away from the antinodes, the distance between floaters increases and the antinode clusters are loosely packed [the clusters surrounded by orange solid lines in Fig. \ref{suppl_Fig3_breathing}(a)]~\cite{Afootnote3}. In addition to Fig. \ref{suppl_Fig3_breathing}, the dynamics of the antinode and node clusters can be observed in Ref. \cite{SupplementaryMovie}.

The breathing mechanism plays a major role in our explanation of why we observe antinode clusters for low $\phi$ and node clusters for high $\phi$. This understanding will be used in Appendix~\ref{sectII} to create artificial floater clusters on a two-dimensional standing wave.

\section{Two-dimensional artificial clusters on a standing wave} \label{sectII}

In this Appendix we explain 
the procedure to artificially create node and antinode clusters incorporating the differences between the two due to the breathing effect for the energy estimation 
discussed in the main text, Section~\ref{est}.

\begin{figure}[h]
\begin{center}
  \includegraphics[width=1.0\linewidth]{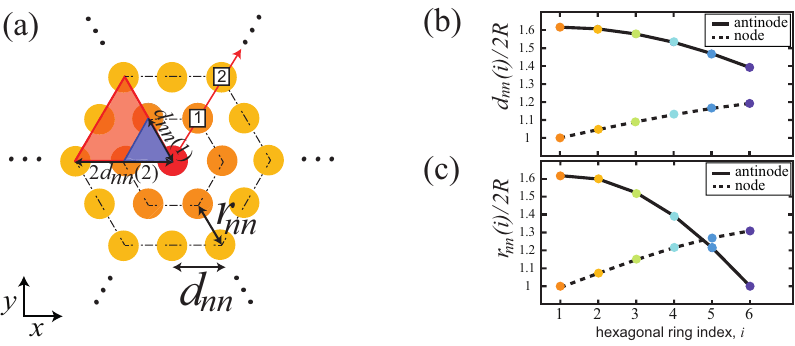}
\end{center}
\vspace{-6mm} \caption{\label{suppl_Fig4_designclusters} \small (Color online) Design of the artificial node and antinode clusters: (a) The center (red) floater, representing the antinode or node, is surrounded by concentric hexagonal rings with index $i$ and marked by dash-dot lines. The nearest-neighbor distance $d_{nn}$ between floater centers within a hexagonal ring and the nearest-neighbor distance between floaters belonging to two consecutive rings $r_{nn}$ are shown. 
(b) The distance $d_{nn}$ is plotted versus the ring index $i$ for both the antinode [solid line] and node clusters [dashed line], where the difference is due to the inclusion of the breathing effect (see text). (c) Same for the distance $r_{nn}$ versus $i$.  
Color coding in (b) and (c) is for illustrative purposes and consistent with that of Fig.~\ref{Fig4_designedclusters}. 
}
\end{figure}

To create the artificial clusters we use monodisperse floaters arranged in a hexagonal packing. We start from a center particle, and to increase $\phi$, hexagonal rings are drawn around this center one as represented in Fig. \ref{suppl_Fig4_designclusters}(a) with dotted dashed lines. The number of floaters in each hexagonal ring is equal to $6i$, where $i$ is the index of each subsequent hexagonal ring.
The difference between the breathing antinode and the non-breathing node clusters is implemented by using a different distance between floaters in consecutive rings. For the antinode clusters we start with a high value of 1.6 times the floater diameter, which decreases for every next ring, whereas for the node clusters we start from a closely packed situation in which the distance is increased with every added ring [see Figs.~\ref{suppl_Fig4_designclusters}(b), \ref{suppl_Fig4_designclusters}(c)].

Quantitatively, we define two distances, namely $d_{nn}$ and $r_{nn}$, which are the distance between the centers of the nearest-neighbor floaters within a hexagonal ring and the distance between the centers of the floaters belonging the nearby hexagonal rings, respectively [see Fig.~\ref{suppl_Fig4_designclusters}(a)].
For the antinode clusters, $d_{nn}(i)$ is defined as
\begin{equation}\label{dnn}
    d_{nn}(i) = 2R + B \cos k\,i2R,
\end{equation} 
where $k$ is the wave number, $R$ the average radius of the floater, and $B$ a length scale of the order of the wave amplitude $a$. From the experimental average distance between the floaters in the antinode clusters [cf. Fig.~\ref{suppl_Fig3_breathing}(a)] we find $B \approx 0.37$ mm. Similarly, for the node clusters we have
\begin{equation}\label{dnn_node}
    d_{nn}(i) = 2R + C \sin k\,i2R,
\end{equation} 
with $C \approx 0.10$ mm, estimated from Fig.~\ref{suppl_Fig3_breathing}(b). 
In Fig.~\ref{suppl_Fig4_designclusters}(b) we show how $d_{nn}(i)$ decreases with $i$ for the antinode cluster and how it increases for the node cluster.

Finally, $r_{nn}(i)$ is derived from a recurrence relation deduced from the equilateral triangles drawn in Fig. \ref{suppl_Fig4_designclusters}(a)]: Each side of the largest equilateral triangle, which extends to the $i$-th ring, is equal to $id_{nn}(i)$. Similarly, each length of the smaller equilateral triangle, extending to the $(i-1)$-th ring, is equal to $(i-1)d_{nn}(i-1)$. Consequently, $r_{nn}$ is given by their difference
\begin{equation}\label{rnn}
     r_{nn}(i)=id_{nn}(i)-(i-1)d_{nn}(i-1)\,.
\end{equation}
The behavior of $r_{nn}$ as a function of $i$ is plotted in Fig.~\ref{suppl_Fig4_designclusters}(c).

\section{Capillary energy and drift energy}\label{sectIII}

In the main text, Sec. \ref{est}, both the capillary energy of each floater pair and the drift energy of a single floater are given without prefactors. 
We now present them with the corresponding prefactors.

First, the capillary energy of two floaters at a distance $l$ in the approximation for small surface deformations is given by~\cite{Capillaryforce}
\begin{equation}\label{capillaryenergy}
    E_c(l) = -2\pi\sigma R^2 B^2 \Sigma^2 K_0(l/l_c),
\end{equation}
where $K_0$ is the zeroth-order modified Bessel function of the second kind and $l_c=\sqrt{\sigma/\rho g}$ is the capillary length. The Bond number $B$ and the dimensionless quantity $\Sigma$ are defined in Appendix~\ref{sect_contact_angle}. 

Secondly, the drift energy $E_d(x,y)$ of a floater located at position $(x,y)$ is calculated by generalizing the one-dimensional drift force given in Eq. (\ref{drift_ave_force}) to two dimensions
\begin{eqnarray}\label{2Ddriftforce}
   \textbf{f}(x,y) &=& \frac{M-m_d}{8}a^2\;\left(\omega^2k_x\sin 2k_x x\;[1-\cos 2k_y y]\;\widehat{\textbf{x}}\right. \nonumber\\
   &+& \left. \omega^2k_y\;\sin 2k_y y[1-\cos 2k_x x]\;\widehat{\textbf{y}}\right)\,,
\end{eqnarray} 
where $\omega$ and $a$ are the angular frequency and amplitude of the Faraday wave, $k_x$ and $k_y$ the (usually similar) wave numbers in the $x$ and $y$ direction, and finally $M$ and $m_d$ are the floater mass and the mass of the displaced water.  $\widehat{\textbf{x}}$ and $\widehat{\textbf{y}}$) denote the unit vectors in the x- and y-direction. This force field is conservative and therefore Eq.~(\ref{2Ddriftforce}) can be integrated and provides the drift energy $E_d(x,y)$ of the floater
\begin{equation}
    E_d(x,y)=-\frac{M-m_d}{16}\;a^2\omega^2\;[1-\cos 2k_x x]\,[1-\cos 2k_y y] .
\end{equation}

\section{Experimental details} \label{sectIV}

\subsection{Frequency and amplitude settings to obtain standing Faraday wave} \label{Frequency and amplitude of a standing Faraday wave}

A standing Faraday wave is generated in a vertically vibrated container filled with a fluid. When the fluid layer is vertically oscillated, a parametric instability of the free surface occurs when the oscillating amplitude $a_0$ becomes greater than a critical amplitude $a_c$. The phenomenon was first investigated by Faraday~\cite{Faraday} and associated with his name after that. In our case, the resultant standing surface wave is a subharmonic response of the vertical driving such that the resonance frequency of the standing wave $f=f_0/2$, where $f_0$ is the shaking frequency at the resonance.

The wavelength of the standing water Faraday wave without floaters can be calculated from  
the inviscid dispersion relation~\cite{Kundu} $\omega^2_k=\left[gk+(\sigma/\rho_{w} k^3)\right]\tanh kH$, where $\omega=2\pi f$, $k$ is the wave number, $g$ is the acceleration of the gravity, and $\sigma$ is the surface tension of the air-water interface.  
In addition, $\rho_{w}$ is the density of the water, and $H$ is the depth of the water. Therefore, the standing Faraday wave with a desired wavelength, matched to the dimensions of the container, can be obtained by adjusting the required $f$ calculated from the dispersion relation.

However, experimentally things are much more complicated than this simple picture suggests: When more than one wavelength fits into the system--as is easily the case in the two-dimensional system that we are using--there are many competing possible modes that are a threat to stability. There are methods to force the system to choose exactly one wavelength with a given $f$ by building a mechanically stable experimental setup~\cite{BoschPhDThesis} or by deforming the sidewalls of a square container~\cite{JDCrawford_nonsquarecontainer}, but these are beyond the scope of this paper. 

Furthermore, in this study, the presence of the floaters alters the physical properties of the water such as $\sigma$ and $\rho_{w}$ near the free surface, and with that also the dispersion relation. Therefore, the resonance frequency $f$ of the wave changes when adding new floaters to the system. As a result, even if a mechanically stable setup would be designed, this would need to be redesigned for each floater concentration $\phi$ due to the varying of $f$. 
Consequently, to simplify the required experimental work to obtain a stationary long-time standing Faraday wave, we apply the following procedure.

We sweep the shaking frequency $f_0$ and amplitude $a_0$ to obtain a rectangular wave pattern with a wavelength that is approximately 30 times larger than the average particle diameter~\cite{Afootnote4}. The resonance frequency is now $f=f_0/2$ and the amplitude of the wave is given by $a$. Subsequently, after adding more floaters to the system, the sweeping procedure is repeated to again find a stable rectangular wave pattern, now for a (slightly) different value of $f=f_0/2$. Also the shaking amplitude $a_0$ needs to be adjusted, both to obtain a well-defined rectangular wave pattern and to obtain a similar wave amplitude $a$.

\begin{figure}[ht]
\begin{center}
  \includegraphics[width=1.0\linewidth]{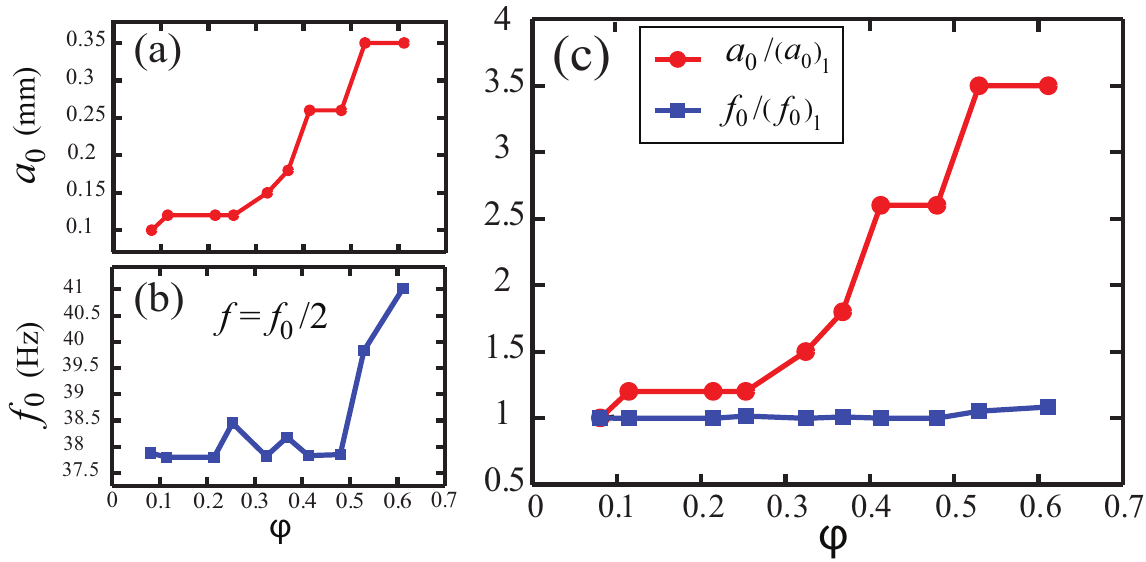}
\end{center}
\vspace{-6mm} \caption{\label{suppl_Fig5_a0f0} \small (Color online) (a) The shaking amplitude $a_0$ and (b) the shaking frequency $f_0$ needed to create a stable standing Faraday wave in the presence of floaters are plotted as a function of the floater concentration $\phi$. The resultant wave pattern is rectangular for each $\phi$ with a wavelength in the range of 17--24 mm both for the $x$ and $y$ directions and frequency $f$ that is half the shaking frequency: $f=f_0/2$. (c) The same data as in (a) and (b) but now normalized by their respective values for the lowest $\phi$. }
\end{figure}

In our experiment, $f_0$ needs to be adjusted in the range of 37--42 Hz, and $a_0$ in the range of 0.10-0.35 mm. 
In Figs.~\ref{suppl_Fig5_a0f0}(a) and~\ref{suppl_Fig5_a0f0}(b), we plot $a_0$ and $f_0$ as a function of $\phi$. From Fig.~\ref{suppl_Fig5_a0f0}(c), where we plot the same data normalized by their values for the lowest $\phi$ it can be appreciated that, whereas $f_0$ only needs to be changed by a few percent, we need to substantially increase $a_0$, namely by a factor $3.5$.\\ 

All of the results presented in this paper 
are obtained for hydrophilic polystyrene particles with a contact angle of $74^\circ$, a density of $1050$ kg/m$^3$, and an average diameter of $0.31$ mm. We have however not restricted ourselves to these particles and tried different types and sizes.
The material and diameter of the spherical particles the behavior of which we have studied on top of the Faraday waves are hydrophobic polystyrene and teflon with a diameter of 3.17 mm, hydrophilic nylon of 1.59 mm, hydrophilic polymethylethylene of 200 $\mu$m, and hydrophilic hollow glass spheres of 30 micron. The largest particles disturb the surface wave to such an extent that it is not possible to obtain a stationary standing wave. On the other hand, small particles are observed to be too mobile and are often entrained into the bulk of the liquid by the surface waves during the experiment. Thus, we have restricted ourselves to the hydrophilic polystyrene spheres described above.

\subsection{Initial conditions}\label{Initial conditions}

On the undisturbed water surface, the floaters initially form clusters (or rafts) induced by the attractive capillary interaction~\cite{Cheerios,Capillaryforce}. To destroy these initial clusters and to produce the stationary standing wave, the method used in Refs.~\cite{Falkovich, FalkovichPRL, FalkovichEurPhys} is applied. An important adjusting parameter in this procedure is $\epsilon=a_0-a_c/a_c$, where $a_c$ is the minimum required shaking amplitude to obtain the parametric instability. First, at a slowly varying frequency, the system is vibrated with a small vibration amplitude $a_0$, namely $\epsilon\ll 1$. Then, $a_0$ is increased considerably such that $\epsilon \gg 1$, to randomize the floater distribution. While keeping a slowly varying frequency, we decrease $a_0$ such that we are still satisfying $\epsilon >1$. When a rectangular stationary standing wave is reached, $a_0$ and $f_0$ are kept fixed. This procedure is repeated after adding more floaters to the system.


%
%
%
%
%


\begin{thebibliography}{apsrev}

\bibitem{Falkovich} G. Falkovich, A. Weinberg, P. Denissenko, and S. Lukaschuk, Nature (London) $\textbf{435}$, 1045 (2005).
\bibitem{FalkovichPRL} P. Denissenko, G. Falkovich, and S. Lukaschuk, Phys. Rev. Lett. \textbf{97}, 244501 (2006).
\bibitem{FalkovichEurPhys} S. Lukaschuk, P. Denissenko, and G. Falkovich, Eur. Phys. J. Special Topics \textbf{145}, 125 (2007).
\bibitem{surfacetension} In this case, both heavy hydrophilic and hydrophobic spheres can float by the help of surface tension.
\bibitem{DominicVella} P. Cicuta and D. Vella, Phys. Rev. Lett. $\textbf{102}$, 138302 (2009).
\bibitem{Cheerios} D. Vella and L. Mahadevan, Am. J. Phys. $\textbf{73}$, 817 (2005).
\bibitem{Capillaryforce} D. Y. C. Chan, J. D. Henry, Jr., and L. R. White, J. Colloid Interface Sci. $\textbf{79}$, 410 (1981).
\bibitem{MichealBerhanu} M. Berhanu and A. Kudrolli, Phys. Rev. Lett. $\textbf{105}$, 098002 (2010).
\bibitem{PlanchetteLorenceauBiance} C. Planchette, E. Lorenceau, and A.-L. Biance, Soft Matter \textbf{8}, 2444 (2012).
\bibitem{Douady} S. Douady, J. Fluid Mech. $\textbf{221}$, 383 (1990).
\bibitem{particlesource} The particles have been custom made and are not commercially available.
\bibitem{cleaningprotocol} C. Duez, C. Ybert, C. Clanet, L. Bocquet, Nature Phys. 3, 180 (2007). See the section Methods in this article for the cleaning protocol of both hydrophilic (wetting) glass material and hydrophobic teflon (non-wetting) particles. Note that we immersed the hydrophilic glass container in piranha solution (1 vol H2O2, 2 vol H2SO4) for 25 min. Since we have hydrophilic polystyrene floaters, the advised heating temperature is $70\,^\circ$C at most. We heated the floaters at $65\,^\circ$C for 1 hour.


\bibitem{spatialaverage} The spatial average is performed over the total horizontal field of view and the time average is computed using the experimental images over 200 wave periods ($\approx200\times0.05=10$ s).

\bibitem{footnote1} In this procedure the patterns are shifted such that the origin coincides with an antinode. To eliminate the contribution of floater motion into the third dimension, only images where the wave elevation is nearly zero are used in the average. 

\bibitem{SupplementaryMovie} See the Supplemental Movie for a video presenting the breathing antinode clusters at low $\phi$ and the non-breathing node clusters at high $\phi$ floating on the standing Faraday wave \url{http://fcxn.files.wordpress.com/2012/04/antinode_node.mov} \href{http://fcxn.files.wordpress.com/2012/04/antinode_node.mov}.




\bibitem{Vassileva} N. D. Vassileva, D. van den Ende, F. Mugele, and J. Mellema, Langmuir $\textbf{21}$, 11190 (2005).


\bibitem{Afootnote} In Refs.~\cite{Falkovich, FalkovichEurPhys}, Eq. (\ref{drift_ave_force}) has a prefactor $1/2$. However, our calculations show that this prefactor should be $1/4$.
\bibitem{Afootnote2} Note that for buoyant spheres ($M<m_d$) the direction of the acceleration reverses in Fig.~\ref{suppl_Fig2_floaterdynamics} and, following the same reasoning, the drift will direct the spheres towards the nodes instead of the antinodes.
\bibitem{Afootnote3} This of course holds for the antinodes that have their maximum during the first half of the wave period. For those that have their minimum in the first half the reverse is true [the clusters surrounded by purple dashed lines in Fig. \ref{suppl_Fig3_breathing}(a)].
\bibitem{Afootnote4} We tested a range of values for the wavelength, namely from 20--40 times larger than the average particle diameter $2R$. However, we successfully obtain long-time stationary standing Faraday waves for each $\phi$ only if the wavelength is 30 times larger than $2R$.
\bibitem{Faraday} M. Faraday, Philos. Trans. R. Soc. London 121, 319 (1831).
\bibitem{Kundu} P. K. Kundu and I. M. Cohen, Fluid Mechanics (Elsevier, Oxford, 2008).
\bibitem{BoschPhDThesis} D. Binks and W. van de Water, Phys. Rev. Lett. \textbf{78}, 4043 (1997).
\bibitem{JDCrawford_nonsquarecontainer} J. D. Crawford, Phys. Rev. Lett. $\textbf{67}$, 441 (1991).

\end{thebibliography}
\end{document}